\documentclass[usenatbib]{mn2e}
\usepackage{graphicx}
\usepackage{epsf}

\def\msun{{\rm M_{\odot}}}
\def\rsun{{\rm R_{\odot}}}

\def\Msun{\hbox{$\rm ~M_{\odot}$}}

\def\H0{{\rm ~km~s^{-1}~Mpc^{-1}}}

\def\msun{M_{\rm \odot}}

\begin{document}

\title[On the origin of black hole spin in Cyg X-1]{On the origin of black hole spin in high-mass black hole binaries: Cygnus X-1}

\author[M. Axelsson et al.]{Magnus Axelsson$^{1}$\thanks{email: magnus.axelsson@astro.lu.se},
Ross P. Church$^{1}$, Melvyn B. Davies$^{1}$, Andrew J. Levan$^{2}$, \newauthor Felix Ryde$^{3}$\\
$^{1}$Lund Observatory, Box 43, SE--221 00, Lund, Sweden.\\
$^{2}$Department of Physics, University of Warwick, Coventry, CV4 7AL \\
$^{3}$Department of Physics, Royal Institute of Technology, AlbaNova, SE-106 91 Stockholm, Sweden
}

\date{Accepted 2010 November 19. Received 2010 November 19; in original form 2010 September 13}

\pagerange{\pageref{firstpage}--\pageref{lastpage}} \pubyear{2002}

\maketitle


\begin{abstract}
To date, there have been several detections of high-mass black hole binaries in both the Milky Way and other galaxies. For some of these, the spin parameter of the black hole has been estimated. As many of these systems are quite tight, a suggested origin of the spin is angular momentum imparted by the synchronous rotation of the black hole progenitor with its binary companion. Using Cygnus X-1, the best studied high-mass black hole binary, we investigate this possibility. We find that such an origin of the spin is not likely, and our results point rather to the spin being the result of processes during the collapse.
\end{abstract}

\section{Introduction}

Observations have so far led to a large number of discovered high-mass X-ray binaries, both in the Milky Way and other galaxies \citep[over 100 in the Milky Way alone;][]{liu06}. However, only a handful of these are believed to harbor black holes. Two of these, Cygnus X-1 and Cygnus X-3, are found in our own Galaxy. The most well-known and well-studied of the black hole binaries is Cygnus X-1. 

In parallel with the discovery of more sources, new techniques have given estimates of the spin parameter of the black hole. The two main techniques rely on spectral fitting of the thermal component \citep[e.g.,][]{sha06} or relativistically broadened iron line \citep[e.g.,][]{mil07}. However, these results are quite sensitive as spectral fitting is prone to a certain amount of degeneracy and model dependency. The technique is also dependent on the spectral state of the source, making it difficult to apply to certain sources which do not show this state, e.g. Cyg~X-1. 

\cite{axe05} presented an alternative technique for determining the spin parameter in Cyg~X-1. By studying the evolution of quasi-periodic oscillations (QPOs), they were able to give support to the idea that the oscillations are connected to the relativistic precession frequencies predicted by general relativity in the strong gravitational field close to the black hole. Based on this identification, the spin parameter was measured to be $a_*=0.48\pm0.01$, assuming the black hole mass to be 9\Msun.
While still dependent on the theoretical interpretation, this method allows for a much more precise determination of the spin than the spectral modelling techniques.

As several of the black hole systems discovered so far are tight systems, a possible source of spin may be tidal locking \citep[see, e.g.,][]{pod04,ldk06}. Upon collapse of the black hole progenitor, the angular momentum is preserved in the spin parameter of the black hole. In this paper, we will investigate whether such a scenario can explain the measured spin parameter of Cyg~X-1. This system is a good candidate for such an investigation: it is bright and thereby well studied, and observations by \cite{mir03} indicate that the mass loss in the formation of the black hole was low.

In the rest of this section, we motivate our choice for the system parameters such as the mass of the compact object. Based on this, in the following section we investigate whether systems similar to Cyg~X-1 can acquire sufficient angular momentum via tidal locking to explain the current spin parameter. In Sect.~\ref{evolution} we discuss the evolution of the system in the past, including possible processes which may cause the spin measured today to differ from the natal one.

\subsection{The mass of the compact object}

Although one of the best studied black hole candidate sources, estimates of the current parameters of the Cyg~X-1 system vary. Perhaps most well-determined is the mass function $f$,

\begin{equation}
f=\frac{\left(M_{\rm co} \sin i\right)^3}{\left( M_{\rm co} + M_2\right)^2} \;,
\end{equation}
where $M_{\rm co}$ and $M_2$ are the masses of the compact object and companion star, respectively, and $i$ the inclination of the system. \citet{gie03} found $f=0.251\pm0.007 \Msun$.   

The mass function is often used to determine the mass of the compact object, and thus requires 
estimates of the inclination and companion mass. These parameters are however difficult to measure. Several different methods have been used to estimate the mass of the donor. \citet{gie86} used the spectroscopic orbit, light curve, photospheric line broadening and an assumed degree of Roche lobe filling to find a mass of $M_2=33\pm 10${\Msun}, giving $M_{\rm co}=16\pm5${\Msun}. From spectroscopic analysis of the line spectrum, \citet{her95} derived mass estimates of $M_2\sim18${\Msun} and $M_{\rm co}\sim10${\Msun}. Observational and evolutionary constraints led \citet{zio05} 
to significantly higher mass estimates: $M_2=40\pm5${\Msun} and $M_{\rm co}=20\pm5${\Msun}.
This wide range of masses illustrates the difficulties inherent in these indirect measurements of
the black hole mass.

Recently, direct determination of the mass of the compact object has been suggested from
measurements of the X-ray radiation. For example, \citet{sha07} study the correlation between low-frequency quasi-periodic oscillations (QPOs) and spectral index and determine the mass of the compact object to $8.7\pm0.8$\Msun, which is a much stronger estimate than achieved by the indirect measurements above.

In this paper, we will use the value of $M_{\rm co}=8-10$ \Msun. This range is in the lower 
end of the values found by studies of the donor star, but nevertheless compatible with both these
measurements and the more direct mass estimates of the compact object. For the inclination, values from $\sim20^\circ$ to $\sim70^\circ$ have been suggested, with most falling in the lower end of the range. We will here use $i=28^\circ$--$38^\circ$, following \citet{gie86}. These values then give us the current system parameters, presented in Table~\ref{partable}.

\begin{table}
\begin{center}
\begin{tabular}{l l l} 
\hline \hline
Parameter & Value & Reference\\
\hline
Mass function ($f$) & 0.251\Msun & \citet{gie03} \\
Black hole mass ($M_{\rm co}$) & 8--10\Msun & See text for details.\\
Inclination ($i$) & 28$^{\circ}$--38$^{\circ}$ & \citet{gie86}\\
Separation ($d_0$) & 32--42 $\rsun$ & From $f$, $M_{\rm co}$ and $i$.\\
Donor mass ($M_{2,0}$) & 7--21 $\Msun$ & From $f$, $M_{\rm co}$ and $i$.\\
Period ($P$) & 5.59982 d & \citet{gie03}\\
Spin parameter ($a_*$) & 0.47--0.49 & \citet{axe05}\\
\hline
\end{tabular}
\end{center}
\caption{Table of parameter values used in this paper. The separation and donor mass ranges are calculated using the mass function with the assumed ranges of black hole mass and inclination.
Note that these ranges are not to be seen as confidence intervals in the traditional sense, but merely
the values possible when $M_{\rm co}$ and $i$ are allowed to vary over our chosen ranges.}
\label{partable}
\end{table}

\section{The orbital history of Cyg~X-1}
\label{evolution}

We now turn to the parameters of the system at the time of collapse. Our starting point is the currently observed system, summarized in Table~\ref{partable}. The measured value of the black hole spin is markedly different from the value inferred assuming tidal locking of the binary with its current orbital parameters. Assuming a symmetric stellar collapse event, this suggests that either the black hole has been spun up since its formation, or alternatively that the binary system was much tighter in the past, providing a higher angular momentum budget for the spinning stellar core at the time of collapse. 

To determine the state of the system at the time of core collapse we evolve Cyg X-1 backwards from its current state. In our backwards extrapolation we assume that the current mass of the black hole ($M_{\rm co} = 9${\Msun}) is equal to the mass of the stellar core which collapsed to create it. This
is consistent with the picture of very low mass loss, less than $\sim1${\Msun}, implied by the low
velocity of the system \citep{mir03}. 

The moment of inertia of the helium core is given by $I = k^2 M_{\rm co} R_{\rm co}^2$.  Before the collapse of the helium core into a black hole, the spin parameter $a_*$ can be expressed
as
\begin{equation}
a_*=\frac{I \Omega_{\rm co} c}{G M_{\rm co}^2}=\frac{k^2 \Omega_{\rm co} R_{\rm co}^2 c}{G M_{\rm co}}\;,
\label{astarexpression}
\end{equation}
where  $\Omega_{\rm co} $ is the rotational frequency and $R_{\rm co}$ the radius of the helium
core. Assuming a polytropic equation of state for the core with index $n=3 $, gives $k=0.275$ \citep{lai93}. The radius of a helium core is given by $R_{\rm co}=0.22 (M_{\rm co}/\msun)^{0.6}\ \rsun$ \citep{lee02}; a 9{\Msun} core has the radius $0.82\; \rsun$. Knowing the spin and mass of the core at the 
time of collapse thus enables us to determine the rotational frequency to be $7.9\times 10^{-5}$ Hz. Under the assumption of tidal locking this is the same as the orbital frequency of the system, corresponding to 
a period of $\sim 0.9$ days. We will now study the evolution of the system to determine the other parameters needed for such an orbit. 

\subsection{Evolution by stellar wind}

The system today is wind-fed, and as a first scenario we consider the evolution of the binary {\em after} the creation of the black hole to be driven by mass loss from the donor star. This mass loss acts to increase the orbital separation while simultaneously decreasing the total mass of the system (assuming the accreted mass can be neglected). This implies that the binary was {\em tighter in the past} than is
{\em measured in the present}. We can estimate this by evolving the binary in time. Assuming the fraction of the mass loss accreted by the compact object to be much less than unity and that the lost mass leaves with specific angular momentum equal to that of the donor, the product of separation $d$ and total mass are constant \citep{ver93} giving 

\begin{equation}
d=\frac{M_{\rm co}+M_{2,0}}{M_{\rm co}+M_2} d_0 \;,
\label{verbunteq}
\end{equation}
where $d$ is the separation and subscript 0 denotes present values. The orbital frequency $\Omega_{\rm orb}$ of the system (assuming Keplerian orbits) is given by

\begin{equation}
\Omega_{\rm orb}=\sqrt{\frac{G (M_{\rm co}+M_2)}{d^3}}=\sqrt{\frac{G(M_{\rm co}+M_2)^4}{d_0^3(M_{\rm co}+M_{2,0})^3}}\;.
\label{omegaexpression}
\end{equation}

Under the assumption of tidal locking, the orbital frequency of the core
is the same as that of the system. Combining Eqs.~\ref{omegaexpression} and 
\ref{astarexpression}, we now get

\begin{equation}
a_*=\frac{c R^2_{\rm co}}{G^{0.5}M_{\rm co}d_0^{1.5}(M_{\rm co}+M_{2,0})^{1.5}}k^2 (M_{\rm co}+M_{2})^2 \;.
\label{ploteqn}
\end{equation}

\begin{figure}
\includegraphics[width=8cm]{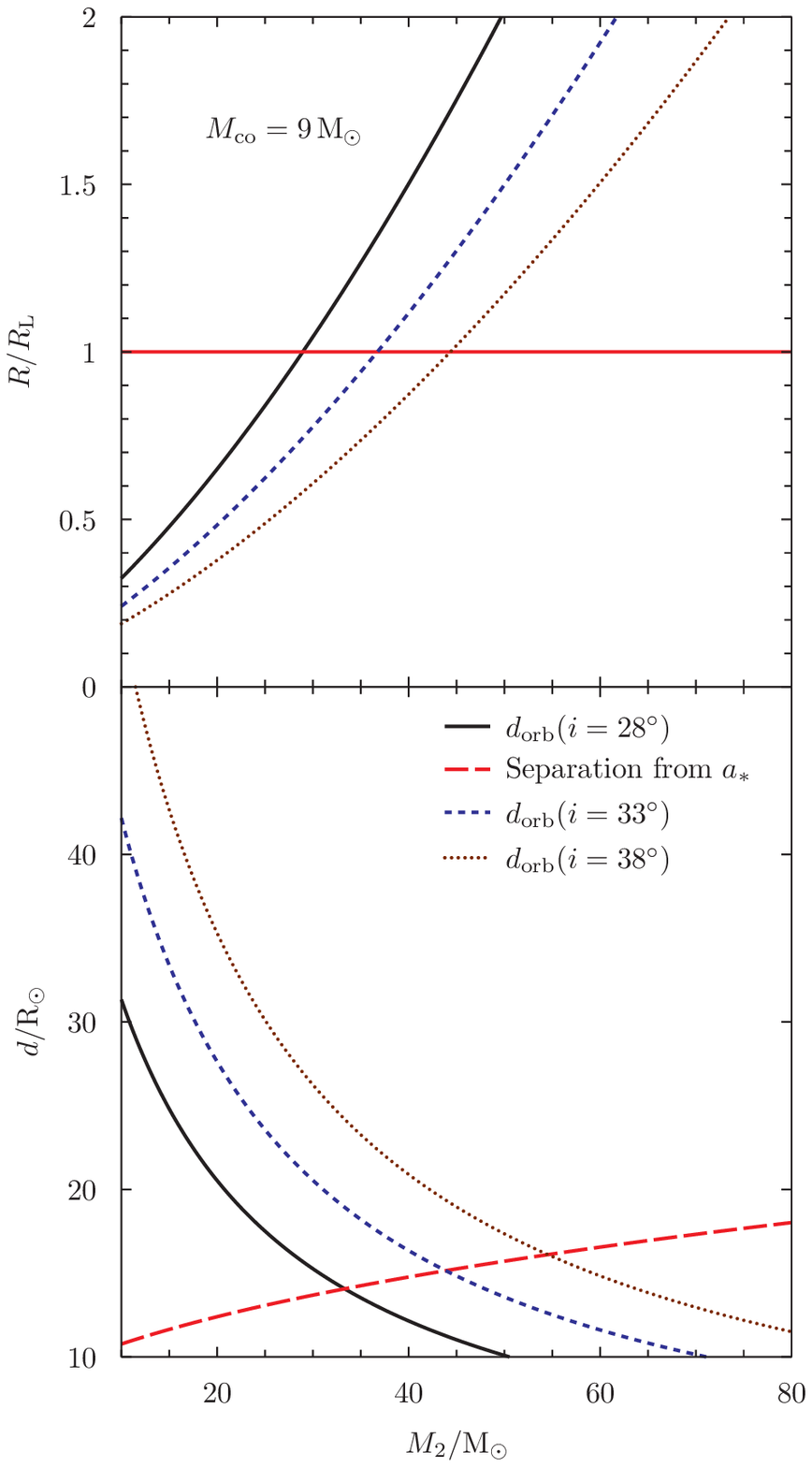}
\caption{Orbital separation (lower panel) and donor radius over Roche lobe (upper panel) as a function of donor mass when the black hole was formed. The long-dashed line in the lower panel indicates the separation required to give the current spin parameter of the black hole found by \citet{axe05}.}
\label{evplot}
\end{figure}

Maintaining the assumption that mass accretion onto the black hole has not
appreciably altered the mass or spin, the spin parameter measured today is that
of the system when the black hole was formed. We can thus use Eq.~\ref{ploteqn}
to determine the mass of the donor star at this time. The lower panel of Fig.~\ref{evplot} shows the
orbital separation as a function of donor mass for our range of inclinations. The long-dashed line is 
the separation required to give the observed value of $a_*$ found by \citet{axe05}. 

To explain the current spin parameter, the mass of the donor must therefore
have been 46{\Msun} at the time the black hole formed, assuming 
$M_{\rm co}=9$~{\Msun} (solid line). 
Equation~\ref{verbunteq} then gives the orbital separation at this time, 
$15\; \rsun$. The dashed lines show the evolutions allowed when including the full range of values in Table~\ref{partable}. Current donor star masses of 7{\Msun} and 21{\Msun} may at black hole formation have been 29{\Msun} and 63{\Msun} respectively, with the orbital separation in these extreme cases being 11 $\rsun$ and 19 $\rsun$.

In order for tidal locking to be viable as the source of spin, the tidal synchronisation timescale must be relatively short. In particular, the tidal locking timescale must be comparable to the widening of the orbit as the companion loses mass. Using this criterion, we estimate the separation at which unlocking occurs by setting the tidal synchronisation timescale equal to $M/\dot M$. We follow the calculations of the synchronisation timescale in section 2.3 of \citet{hur02}, assuming a radiative envelope for the He core. Assuming a mass loss rate of $\sim 10^{-6}$\,{\Msun}~yr$^{-1}$ (see further in Sect.~\ref{spinup} below), the separation at which the core unlocks (and hence the maximum separation at which tidal locking can be effective) is $\sim13\;{\rsun}$. Given the uncertainties in estimating the tidal interactions, this is entirely compatible with the orbital separation required by the spin parameter. 

To judge the feasibility of these results, we study the ratio between the donor Roche lobe and 
its radius (upper panel of Fig.~\ref{evplot}), assuming the donor is on the main sequence. As the donor mass increases, so does its radius, while the orbital separation and Roche lobe radius decrease. For the evolution described above, we see that the donor would fill its Roche lobe before the required orbital separation and mass are reached. In this case, heavy mass transfer would occur from the donor to the compact object, and the result would likely have been a merger rather than the binary required to 
produce the system we observe today.

These calculations were performed assuming a polytropic index of 3, giving a value of $k=0.275$. While such a polytrope accurately describes massive white dwarfs, it is not clear that it correctly matches the stellar core.  As this most likely contains a central, denser Fe core it may be more centrally concentrated than an $n=3$ polytrope. Thus, we expect that the adopted value of $k$ is on the high side - the real value may well be lower. Eqs~\ref{verbunteq} and \ref{ploteqn} show that a lower value of $k$ leads to an even tighter system being required at black hole formation, aggravating the difficulty.

Clearly, merely adding mass to the donor cannot produce the system required at tidal locking. We must therefore look more closely at the evolution prior to the formation of the black hole.

\subsection{Rotational mixing}

In tight systems, the rapid rotation of the stars can lead to rotational mixing. \citet{demink09} recently showed that this can lead to chemicaly homogeneous evolution of the stars, and the stellar radius will then be much smaller than predicted by normal stellar evolution. While they have yet to model a system such as the ones described above, \citet{demink09} point to a new channel of binary evolution where Roche lobe overflow and subsequent mass transfer is significanty delayed or avoided althogether due to rotational mixing \citep[Case M in][]{demink09}. Interestingly, tight binary systems with massive stars have been discovered, e.g. CQ Cep. In this system the stellar masses are 24$\msun$ and 30$\msun$, and the orbital period is 1.6d. One might therefore speculate whether Cygnus X-1 is the result of such an evolutionary channel. 

In the previous section we derived the required parameters when the black hole was formed. At this
time, the system consisted of a helium core of $\sim9${\Msun} orbiting a stellar companion of 
$\sim 46${\Msun}. The orbital separation was $15\; \rsun$, and the orbital period 0.95 days. The 
separation can be compared to the Roche lobe of the donor, which was $8\;\rsun$. This is somewhat
smaller than the predicted radius of a zero-age main sequence (ZAMS) 46 M$_{\odot}$ star (which is $\sim 10\;\rsun$), and comparable to the radius of a star in the lower end of our derived range.

From these results it is clear that even at ZAMS, the donor would have been too large. While rotational mixing may keep a star from expanding as it evolves, it does not reduce the size at ZAMS and cannot be invoked as a means of enabling the tidal locking required.

\subsection{Different evolutionary scenarios}

In order to allow for a wider orbit, and slightly less massive companion, one may investigate the results if a slightly more massive He core at tidal locking is assumed. The separation is a slow function of mass, so the orbit does not widen much. However, the Roche lobe radius is a slightly stronger function of mass so the Roche lobe radius of the companion actually shrinks slightly even though the orbit widens. It is thereby clear that a more massive He core will not allow for tidal locking at the required frequency.

Another alternative to consider is that the system forms with two main sequence stars close in mass. The system undergoes common envelope evolution, leading to two close He stars at which point tidal locking occurs. Winds then widen the binary to present parameters. The scenario is attractive as the secondary does show chemical peculiarities and is enriched in He. However, with a mass ratio close to unity the evolution of the two stars would be very similar, and by the time the primary becomes a black hole the secondary would have lost too much mass to represent the current system.

\subsection{Possible spin-up?}
\label{spinup}

So far we have neglected any effect of accreted mass and angular momentum on the evolution of the system, and assumed that the spin is solely determined by the system parameters at tidal locking. We can however estimate the possible change in spin contributed by accretion onto the black hole. The current mass loss rate from the donor has been estimated by \citet{gie03} to $\dot{M}_{wind} = 3 \times 10^{-6}$ M$_{\odot}$ yr$^{-1}$. \citet{vri07} find a higher mass loss rate, $\dot{M}_{wind} = 5 \times 10^{-6}$ M$_{\odot}$ yr$^{-1}$. The mass accretion rate is clearly variable; however, following 
\citet{vri07}, we can estimate it for a relatively bright state ($L_X = 10^{38}$ ergs s$^{-1}$). 
In this case $\dot{M}_{capture} = L_X  / e c^2$, where $e=0.42$ for a 
maximally rotating black hole \citep{st83}. 
This yields $\dot{M}_{capture} \sim 5 \times 10^{-9}$ M$_{\odot}$ yr$^{-1}$. For Cyg~X-1, a more
realistic value is $e=0.1$, giving $\dot{M}_{capture} \sim 10^{-8}$ M$_{\odot}$ yr$^{-1}$. The age of the system is difficult to determine, but \citet{mir03} estimated it to be $\sim7 \times 10^6$ years by comparing the relative velocity and distance between Cyg X-1 and its assumed formation site in the Cygnus OB3 association of massive stars. The black hole mass could thus only have increased by $\sim0.1{\Msun}$ in the time since formation.
 
In order to significantly change the spin, the black hole must have accreted $>1$ {\Msun}. For a 10 {\Msun} black hole, the Eddington accretion rate is $\sim 10^{-7}$ {\Msun} yr$^{-1}$. Thus, the black hole would need to accrete at the Eddington rate for $\sim10^7$ years to significantly affect the spin. This is not a likely scenario. In addition, the current accretion rate is merely a few per cent of the Eddington rate. Therefore, the accreted mass can be neglected to first order. The low mass accretion onto the black hole also means that the spin parameter cannot have been changed by mass transfer from the companion star. $\Delta a_* \leq 0.02$ for an accreted mass of 0.1{\Msun}, hence the value measured now is the same as when the black hole was formed. 

Effects such as beaming may lead us to underestimate the luminosity, and thereby the accretion rate. However, studies of X-ray binaries show that the transition 
between the so-called hard and soft states typically occur around a few per cent of the Eddington luminosity \citep{done07}. Spectral studies show that Cyg~X-1 is mostly observed in the hard state; 
thus, the accretion rate cannot be very high even if beaming leads us to underestimate it. Another possibility which may lead us to underestimate the accretion rate is asymmetric accretion. In this case, the radiative efficiency can be much lower than during spherical accretion. While such a scenario is unlikely at present, we cannot rule out that the system has undergone a period of
asymmetric accretion in the past, and thereby accreted more mass than the estimate above. However, the accretion rate is still limited by the capture rate of the stellar wind from the companion,
and we find it unlikely that asymmetric accretion could have increased the accreted mass to the level required. One possibility to do so is to invoke a period of super-Eddington accretion
after the black hole has formed. Such a period would have to be very short, and the accretion rate therefore extremely high. While this scenario cannot be ruled out, the feasibility of
super-Eddington accretion is still under debate.

It is possible that the system underwent a period of mass transfer before the creation of the black hole, and this may have spun up the stellar core. We will therefore consider the case of mass accretion onto the helium core. Although mass transfer may have occurred also at earlier stages in the binary evolution, accretion at this stage will primarily spin up the envelope of the star, and it is unclear how much angular momentum is actually gained by the core.

The maximum time during which the He core can accrete matter is determined by its evolution, and is of the order of $10^6$ years. In order to increase the rotational frequency sufficiently to change the spin parameter by $\Delta a_* = 0.1$, the core would have to accrete more than 1{\Msun} (assuming a radius of $R_{\rm co}=0.8R_{\odot}$).

\section{Other systems}

In this context it is also interersting to note that many high mass black hole binaries discovered so far have very short orbital periods. For comparison, we summarize their parameters in Table~\ref{hmbhbtable}.

\begin{table}
\begin{center}
\begin{tabular}{l l l l l} 
\hline \hline
System & $M_{\rm BH}$ & $M_{\rm comp}$ & $P_{\rm orb}$ & Reference\\
 {} & ($\msun$) & ($\msun$) & {} & \\
\hline
Cyg X-1 & 9 & 14 & 5.6d & see Table~\ref{partable}\\
Cyg X-3 & $\sim30$ & $<60$ & 4.8h & 1 \\
LMC X-1 & 11& 32 & 3.9d & 2 \\
LMC X-3 & 12 & 40 & 1.7d & 3 \\
M33 X-7 & 16 & 70 & 3.5d & 4 \\
IC 10 X-1 & 28 & 35 & 1.45d & 5 \\
NGC 300 X-1 & 17 & 21 & 1.35d & 6 \\
\hline
\end{tabular}
\end{center}
\caption{Parameters of known HMBHB systems. The nature of the compact object in Cyg~X-3 is still unclear, but the currently favored scenario is that of a black hole. References: (1) \citet{hja09}, (2) \citet{oro09}, (3) \citet{yao05}, (4) \citet{oro07}, (5) \citet{pres07}, (6) \citet{crow10}.}
\label{hmbhbtable}
\end{table}

From these values it is clear that Cyg~X-1 is not an unusually tight system, nor is the spin parameter particularly high. The only other high-mass black hole binaries with spin estimates are LMC~X-1 and M33~X-7, but in these cases the spin parameter is estimated from spectral fitting to be $a_*\sim 0.9$ \citep{gou09} and $a_*\sim 0.8$ \citep{liu08}, respectively, which is much higher than in Cyg~X-1. Such a high spin is highly unlikely to arise due to tidal locking alone. Rather, it would appear that explaining the spin parameter in many of the other systems is an even greater challenge! The fact that all measured spins are quite high reduces the probability for a very peculiar evolutionary channel. In our view, it strengthens the case for a high spin being created during the stellar collapse, e.g. through asymmetric accretion. For example, it has been shown that a spherical accretion shock instability (SASI) can form in supernovae, leading to accretion of significant angular momentum \citep{blo06}.
  
\section{Summary and Conclusions}

We have investigated the possibility for the spin of Cygnus X-1 to result from the collapse of a stellar core, rapidly spinning through tidal locking. We find that such a scenario is not compatible with standard stellar evolution models; the required orbit is so close that the companion star would not fit at ZAMS. Rotational mixing or alternative evolutionary paths do not offer a solution, nor is the black hole likely to have been spun up due to accretion. Our conclusion is thus that the spin originates in the black hole formation event.

\section*{Acknowledgements}
We thank Milan Battelino, Stefan Larsson, and Kjell Rosquist for interesting discussions. This work was supported by The Swedish Research Council (grants 2009-691 and 2008-4089) and the Swedish National Space Board. RPC acknowledges support from the Wenner-Gren Foundations. AJL is grateful to the STFC for fellowship support.

\end{document}